\documentclass[twocolumn]{aastex631}

\usepackage{soul} 
\usepackage{amsmath}
\usepackage{xspace}
\usepackage{xifthen}
\usepackage{eso-pic}

\definecolor{forestgreen}{HTML}{228B22}
\definecolor{urlblue}{HTML}{000000}

\mathchardef\mhyphen="2D

\newlength{\dhatheight}

\newcommand{\unit}[1]{\ensuremath{\mathrm{\,#1}}\xspace}

\newcommand{\Gyr}{\unit{Gyr}}

\newcommand{\kpc}{\unit{kpc}}

\newcommand{\e}{\unit{e^{-}}}

\newcommand{\secref}[1]{Section~\ref{sec:#1}}

\newcommand{\figref}[1]{Figure~\ref{fig:#1}}

\newcommand{\bandvar}[2][]{%
  \ifthenelse{\isempty{#1}}{\var{#2}}{\var{#2\_#1}}%
}

\newcommand{\var}[1]{\ensuremath{\texttt{\MakeUppercase{#1}}}\xspace}

\providecommand\physrep{\ref@jnl{Phys.~Rep.}}%
\providecommand\apjs{\ref@jnl{ApJS}}%
\providecommand{\jcap}{\ref@jnl{JCAP}}%

\begin{document}

\title{The DECam Field of Streams: a deep view of the Milky Way halo}

\author[0000-0001-6957-1627]{Peter Ferguson}
\affiliation{DiRAC Institute, Department of Astronomy, University of Washington, 3910 15th Ave NE, Seattle, WA, 98195, USA}

\author[0000-0003-2497-091X]{Nora Shipp}
\affiliation{DiRAC Institute, Department of Astronomy, University of Washington, 3910 15th Ave NE, Seattle, WA, 98195, USA}

\email{pferguso@uw.edu, nshipp@uw.edu}

\begin{abstract}

We present a Field of Streams visualizing stellar structures in the Milky Way halo as viewed by the Dark Energy Camera (DECam). We use $g$- and $r$-band imaging from the Dark Energy Survey and the DECam Legacy Survey, covering $18{,}700 \deg^2$ across the sky. Using an isochrone-based matched filter in $g$ vs. $g-r$, we select old and metal-poor stars in three distance bins, and generate a false-color RGB image of the number density of selected stars.
The DECam Field of Streams shows a variety of Milky Way halo structures, including dwarf galaxies, globular clusters, and an abundance of stellar streams, illustrating the significant progress that has been made in recent years in uncovering the building blocks of the Milky Way's stellar halo in deep, wide-area photometric surveys. This view of our Galaxy will be improved in the coming years as the Vera C. Rubin Observatory Legacy Survey of Space and Time (LSST) begins to collect data to greater depths and across a larger fraction of the sky than ever before.
\end{abstract}


\section{Introduction}
\label{sec:intro}
Wide-area astronomical surveys have transformed our view of the Milky Way's stellar halo, revealing faint stellar systems at all stages of tidal disruption. The Field of Streams in the Sloan Digital Sky Survey \citep{Belokurov:2006} provided a first glimpse at the diversity of tidal debris in the Galactic halo, ranging from massive dwarf galaxy streams (Sagittarius and Orphan-Chenab) to the thin tidal tails extending from globular clusters (Pal 5 and NGC 5466). Over the past two decades, additional surveys have further revealed the population of Milky Way stellar streams \citep[e.g.][]{Bernard:2016, Shipp:2018, Malhan:2018}, with the total number of candidates now surpassing 100 \citep{Mateu:2023, Bonaca:2024}.

These data have enabled a broad range of science pertaining to the formation and structure of the Milky Way, the physics of small-scale galaxy formation, and the distribution of dark matter in the Local Universe. For example, stellar streams have provided measurements of the mass of our Galaxy and the dynamical impact of its most massive satellite, the LMC \citep{Erkal:2019, Shipp:2021, Vasiliev:2021, Koposov:2023}; they have revealed hints of dark substructure in the Milky Way \citep{Price-Whelan:2018, Bonaca:2019, Li:2021, Hilmi:2024}; and they have been used to piece together the Galactic accretion history \citep[e.g.][]{Bonaca:2021, Li:2022}. These impactful scientific results have been possible only through the systematic observations of stellar streams in large astronomical surveys.

\begin{figure*}
    \centering
    \includegraphics[width=0.9\linewidth]{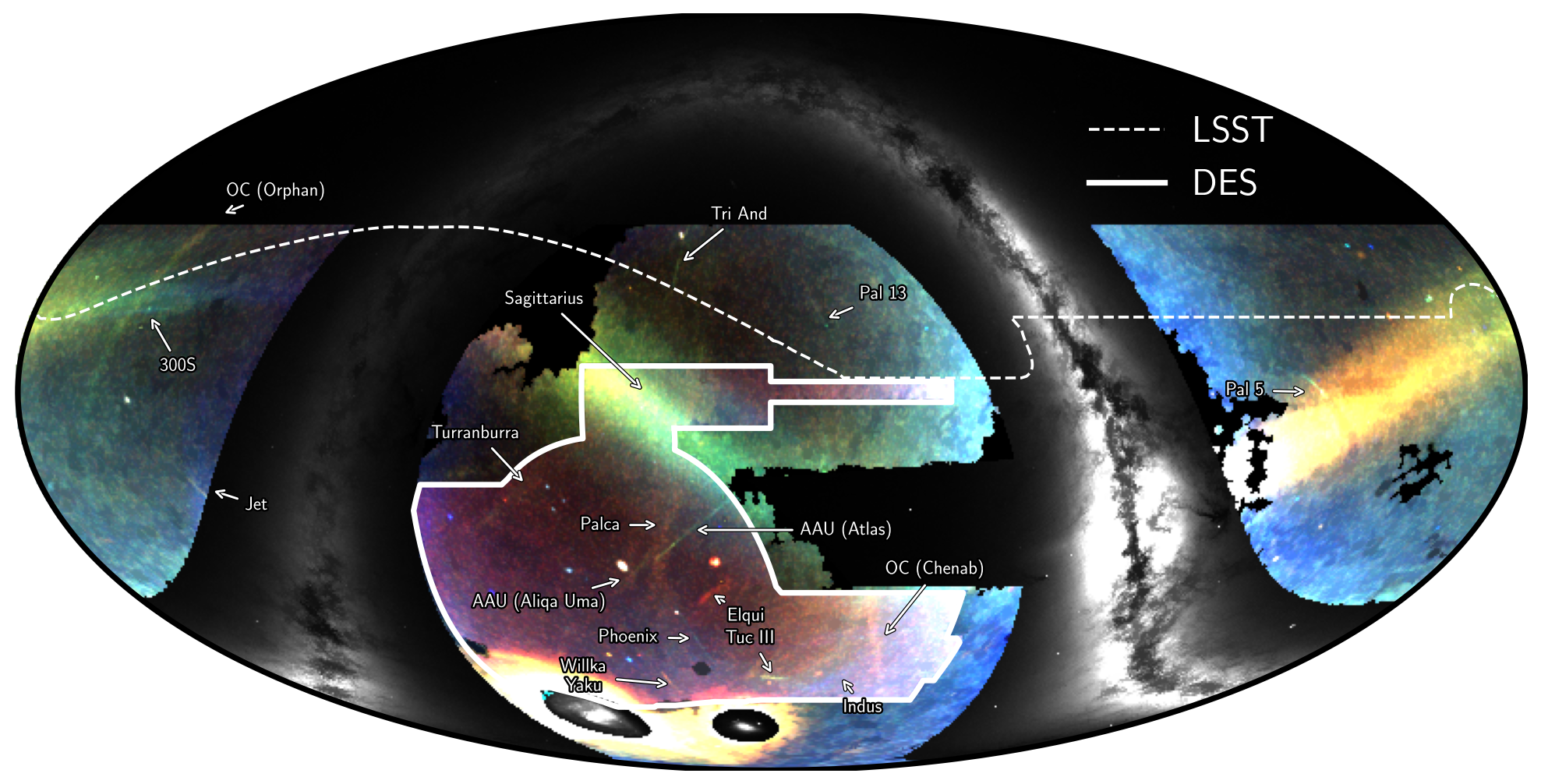}
    \caption{The DECam Field of Streams shows the spatial density of stars that are consistent with belonging to a metal poor stellar population in color-magnitude space. The RGB color channels correspond to the filter applied at three different distance bins with blue showing the closest bin ($10 < d_\odot<15$) \kpc, Green an intermediate bin ($15 < d_\odot < 26$ \kpc) and red the furthest bin ($26 < d_\odot < 100$ \kpc). This representation of the data highlights a large amount of substructure in the Milky Way halo. A subset of the most prominent stellar streams are labeled and are described in \secref{fos} and the point/circular sources seen in the map are resolved dwarf galaxies and star clusters. Additionally, we show the DES footprint as a solid white line and the upper edge of the expected LSST footprint as a dashed white line. See \url{https://github.com/psferguson/decam_fos} for versions of this figure without annotations.}
    \label{fig:fos}
\end{figure*}

Here, we present an updated ``Field of Streams'' as viewed by the Dark Energy Camera (DECam; \figref{fos}). In \secref{data} we describe the data and analysis method and in \secref{fos} we present the DECam view of tidal debris structures in the Milky Way's stellar halo. 
Additionally, we provide the code, and a minimal set of data, used to generate the DECam Field of Streams on GitHub.\footnote{See \url{https://github.com/psferguson/decam_fos}}

\section{Data}
\label{sec:data}
The data presented here comes from three different surveys -- the Dark Energy Survey (DES), the DECam Legacy Survey (DECaLS), and \emph{Gaia}.
Both DES and DECaLS use the data taken by the Dark Energy Camera \citep[DECam;][]{Flaugher:2015} using the 4m Blanco telescope at the Cerro Tololo Inter-American Observatory (CTIO). 

The DES data is taken from DES Y6 Gold \citep{Bechtol:2025} release of DES \citep{Abbott:2021} processed with the DESDM pipeline \citep{Morganson:2018}. 
This data covers ${\sim} 5{,}000 \, \mathrm{deg}^2$ in the southern Galactic cap.  
For this analysis, we select stellar sources with high quality observations, $\texttt{EXT\_MASH} \le 1$, $\texttt{FLAGS\_GOLD} == 0$, below a faint magnitude limit of $\texttt{SOF\_PSF\_MAG\_G} \le 24.0$. 
Magnitude measurements are taken from the \texttt{SOF\_PSF\_MAG\_{BAND}} columns and extinction corrected, using the re-normalized \citep{Schlafly:2011} E(B-V) maps from \citet{Schlegel:1998}, following Section 4.2 of \citet{Abbott:2018}.

For the rest of the southern sky ($\delta  < 33$ deg \& $|b| \gtrsim 20$) we use data taken from the tenth data release of DECaLS \citep[][]{Dey:2019}, processed using the \emph{Tractor} \citep{Lang:2016}. 
This survey covers an additional ${\sim}13{,}700 \, \mathrm{deg}^2$ to a shallower depth. 
To select a pure sample of stars in this survey, we select high quality sources (\texttt{ANYMASK} == 0, \texttt{FRACFLUX} $< 0.05$) that are classified as stars ($\texttt{TYPE} == \texttt{PSF}$). 
We again correct the observations using the SFD98 maps and place a faint magnitude limit of of $g \le 23.5$.
Subsequent references to $g$ and $r$-band magnitudes for DES and DECaLS refer to these extinction corrected values. 

Finally, in the region of the Milky Way plane ($|b| < 20$ deg) and around the MCs ($|R_{\rm LMC}| < 6.5$ deg, $|R_{\rm SMC}| < 5$ deg) we replace the matched filter map with a stellar density map of all stars from \emph{Gaia} DR3.

\subsection{Matched filter selection}
\label{sec:filter}

To highlight faint stellar structures in the Milky Way's halo, we select old and metal-poor stellar populations using an isochrone-based matched filter. We build a filter based on isochrones with an age of $12 \Gyr$ and metallicity $Z=0.0001$ ($\mathrm{[Fe/H]}=-2.17$) from the stellar evolution models of \citet{Dotter:2016}. The matched filter shape in $g$ and $g-r$ space is parametrized as in \citet{Shipp:2018}, and the parameters are selected to empirically match globular clusters and low-mass dwarf galaxy stellar populations in the data. In particular, the width of the selection includes twice the median $g$-band magnitude photometric error, plus an asymmetric color spread that is constant with magnitude, $C = (0.05, 0.1)$, as well as a distance modulus spread of $\Delta (m-M) = 0.5$.

We shift this filter in apparent magnitude to select old, metal-poor stars at a range of distances throughout the stellar halo. We select stellar populations at $m-M = 15 - 20$ in increments of 0.1 mag, corresponding to distances of 10 to 100 \kpc. We then bin the selected stars at each distance modulus into equal area on-sky healpix, with $n_{\rm side} = 512$. To create the false-RGB image in \figref{fos}, we combine the on-sky maps within three different distance slices ($<15,\ 15 - 26,\ \mathrm{and}\ >26$ \kpc), by averaging the stellar density in corresponding healpix across the slices. In the image, the most distant bin is assigned to be red and the closest bin is assigned to be blue.

We also adjust the visualization to account for variations in survey depth. The DECaLS data is made up of archival observations, resulting in large spatial variations in the depth of the catalog. Regions with greater depth have less contamination from faint sources and thus fewer total sources make it into the matched-filter selection. 
To account for this effect, we use survey depth maps (\texttt{PSFDEPTH} for $g$ and $r$-bands) to select regions of increased depth and normalize them with an additive bias so that they have the same average number of counts as the shallower portions of the catalog.
Then, we mask out regions with non-contiguous coverage and areas of low galactic latitude ($|b| < 20$) that are dominated by contamination from the Milky Way disk. 
Finally, to reduce the impact of small gaps in the data we perform a diameter closing operation from \texttt{skimage}.


\section{DECam View of the Stellar Halo}
\label{sec:fos}

\figref{fos} presents the DECam Field of Streams, revealing deep, uniform observations of tidal debris structures across the entire southern sky and out to distances of $100 \kpc$.

As described above, the figure shows the density of old and metal-poor stellar populations colored by distance, where blue represents the populations that are closest to us ($10 < d < 15 \kpc$), green represents intermediate distances ($15 < d < 26 \kpc$), and red highlights populations that are farthest away ($26 < d < 100$ \kpc). 
In addition, we mask out low-Galactic latitude regions and around the MCs since contamination from the Milky Way and the MCs would saturate the matched filter maps. Instead, we replace these regions with a normalized version of the total density of stars in \textit{Gaia} DR3 for visualization purposes. 
We outline the DES footprint on the sky (solid white line), since the region stands out due to the difference in data quality.
This difference is due to the homogeneous way in which data was taken for DES, and the extensive efforts to obtain a precise photometric calibration in this region through the Forward Global Calibration Model (FGCM; \citealt{Burke:2018}) 

Several distinct features stand out in the Field of Streams and are annotated in \figref{fos}. The Sagittarius stellar stream is the bright band that stretches across the projection, as it wraps around the Milky Way multiple times. The LMC and SMC are also bright features, located at the southern edge of the DES footprint. Within the DES footprint, the stellar streams observed by \citet{Shipp:2018} are visible, including ATLAS \citep[first discovered by][]{Koposov:2014} and Aliqa Uma (two components of a single perturbed stream, near the center for the DES footprint), Elqui (the most distant DES stream at $50 \kpc$ shows up as a red streak in the Field of Streams), Tucana III \citep{Drlica-Wagner:2015}, Phoenix \citep{Balbinot:2016}, and in the lower-right part of the DES footprint, Chenab extends northward, ultimately connecting to the Orphan stream, which was discovered in early SDSS data \citep{Belokurov:2006, Grillmair:2006}. Other streams can be seen in other parts of the footprint, including the Jet stream \citep{Jethwa:2018}, the Triangulum-Pisces stream \citep{Bonaca:2012}, the 300S stream \citep{Geha:2009}, and the tidal tails extending from globular clusters Palomar 5 \citep{Odenkirchen:2001} and Palomar 13 \citep{Shipp:2020}.
In the near future, LSST will provide an even more complete census of the population of stellar streams around the Milky Way, extending these observations across half the sky and to the outer limits of the Milky Way.


\begin{acknowledgments}
\end{acknowledgments}

\vspace{5mm}

\bibliography{main}{}
\bibliographystyle{aasjournal}

\end{document}